\documentclass[12pt]{article}

\usepackage{latexsym} 
\usepackage{amssymb}  
\usepackage{amsbsy}   
\usepackage{graphicx}       
\usepackage[dvips]{color}
\usepackage{bm}   

\newcommand{\al}{\alpha}
\newcommand{\be}{\beta}
\newcommand{\ga}{\gamma}
\newcommand{\Ga}{\Gamma}
\newcommand{\de}{\delta}
\newcommand{\De}{\Delta}
\newcommand{\ep}{\varepsilon}

\newcommand{\la}{\lambda}
\newcommand{\La}{\Lambda}

\newcommand{\Si}{\Sigma}
\renewcommand{\th}{\theta}   

\newcommand{\om}{\omega}
\newcommand{\Om}{\Omega}

\newcommand{\beq}{\begin{equation}}
\newcommand{\eeq}{\end{equation}}
\newcommand{\ba}{\begin{array}}
\newcommand{\ea}{\end{array}}
\newcommand{\bea}{\begin{eqnarray}}
\newcommand{\eea}{\end{eqnarray}}
\newcommand{\bi}{\begin{itemize}}  
\newcommand{\ei}{\end{itemize}}
\newcommand{\ben}{\begin{enumerate}} 
\newcommand{\een}{\end{enumerate}}
\newcommand{\bc}{\begin{center}}
\newcommand{\ec}{\end{center}}

\newcommand{\p}{\partial}
 
\newcommand{\txt}{\textstyle}
\newcommand{\dsp}{\displaystyle}
\newcommand{\ns}{\normalsize}
\newcommand\eqn[1]{(\ref{#1})}      

\newcommand{\half} {{\txt \frac{1}{2}}}

\newcommand{\quarter}{{\txt\frac{1}{4}}}

\newcommand{\sixth}{{\txt \frac{1}{6}}}

\newcommand{\tr}{\mbox{tr}}
\newcommand{\Tr}{\mbox{Tr}}

\newcommand{\nn}{\nonumber \\}

\newcommand{\MeV}{{\rm MeV}}

\newcommand{\dmu}{\delta\mu}

\newcommand{\muKp}{{\mu_{K^+}}}

\newcommand{\mueff}{\mu^{\rm eff}}
\newcommand{\gaeff}{\ga_{\rm eff}}
\newcommand{\dm}{\delta m}
\newcommand{\mueffKz}{{\mu^{\rm eff}_{K^0}}}
\newcommand{\mueffKp}{{\mu^{\rm eff}_{K^+}}}
\newcommand{\pbar}{{\bar{p}}}
\newcommand\deriv[2]{{\frac{\partial #1}{\partial #2}}}

\usepackage[normalem]{ulem}  

\newcommand{\feyn}[1]{
  \setbox0=\hbox{\ensuremath{#1}}
  \hbox to\wd0{\hbox to0pt{\hbox to\wd0{\hss/\hss}\hss}\box0}}

\makeatletter 

\def\appendix{\par                              
    \setcounter{section}{0}                     
    \setcounter{subsection}{0}
    \renewcommand{\theequation}{\Alph{section}.\arabic{equation}}
    \renewcommand{\thesection}{Appendix \Alph{section}
                \setcounter{equation}{0}  } 
    \renewcommand{\thesubsection}{\Alph{section}.\arabic{subsection}}
}

\def\applabel#1{\@bsphack
  \protected@write\@auxout{}%
         {\string\newlabel{#1}{{\Alph{section}}{\thepage}}}%
  \@esphack}

\def\section{
\setcounter{equation}{0}        
\@startsection {section}{1}{\z@}{-3.5ex plus -1ex minus 
 -.2ex}{2.3ex plus .2ex}{\large\bf}}
\renewcommand{\theequation}{\arabic{section}.\arabic{equation}}

\def\subsection{\@startsection{subsection}{2}{\z@}{-3.25ex plus -1ex minus 
 -.2ex}{1.5ex plus .2ex}{\normalsize\bf}}

\def\subsubsection{\@startsection{subsubsection}{3}{\z@}{-3.25ex plus
 -1ex minus -.2ex}{1.5ex plus .2ex}{\normalsize\it}}

\makeatother   

\begin{document}

 
\title{\bf Bulk viscosity due to kaons\\ in color-flavor-locked quark matter}

\author{
Mark G.~Alford and Matt Braby\\
\ns Department of Physics \\ \ns Washington University \\
\ns St.~Louis, MO~63130 \\ USA \\[2ex]
Sanjay Reddy \\
\ns Theoretical Division \\
\ns Los Alamos National Laboratory \\
\ns Los Alamos, NM~87545\\ USA \\[2ex]
Thomas Sch\"afer \\
\ns Physics Department \\
\ns North Carolina State University \\
\ns Raleigh, NC 27695\\ USA \\[2ex]
}

\date{24 Jan 2007\\
Revised 20 Aug 2007 \\[1ex]
LA-UR-07-0429}

\begin{titlepage}
\maketitle
\renewcommand{\thepage}{}          

\begin{abstract}
We calculate the bulk viscosity of color-superconducting
quark matter in the color-flavor-locked (CFL) phase. We assume that
the lightest bosons are the superfluid mode $H$ and the kaons
$K^0$ and $K^+$, and that there is no kaon condensate.
We calculate the rate of strangeness-equilibrating processes
that convert kaons into superfluid modes, and the resultant
bulk viscosity. We find that for oscillations with a
timescale of milliseconds, at temperatures $T\ll 1$~MeV,
the CFL bulk viscosity is much less than that of unpaired quark matter, 
but at higher temperatures the bulk viscosity of CFL 
matter can become larger.
\end{abstract}

\end{titlepage}


\section{Introduction}
In this paper we calculate the bulk viscosity of
matter at high baryon-number density (well above nuclear density) and 
low temperature (of order 10 MeV). On theoretical grounds
it is expected that, at
sufficiently high baryon-number density, 
three-flavor matter will be accurately described as a
degenerate Fermi liquid of weakly-interacting quarks, with
Cooper pairing at the Fermi surface (color superconductivity \cite{Reviews})
in the color-flavor-locked (CFL) channel \cite{CFL}. That is the
phase that we will study---the 
phase diagram at lower densities remains uncertain.
The motivation for this calculation is that
in nature the highest baryon-number densities are attained
in the cores of compact stars, and it is speculated that
quark matter, perhaps in the CFL phase, may occur there. This means that
our best chance of learning about the high-density region of the phase
diagram of matter is to make some connection between the physical properties of
the various postulated phases of dense matter and the observable behavior
of compact stars. Calculating the bulk viscosity of CFL quark matter
is part of that enterprise.

Current observations of compact stars are able to give us measurements of
quantities such as the mass, approximate size, temperature, spin and
spin-down rate of these objects. These estimates are steadily improving,
and other quantities, such as X-ray emission spectra, are becoming
available. However, it is a challenge to connect these distantly-observable
features to properties of the inner core of the star, where quark
matter is most likely to occur
(for reviews, see Ref.~\cite{Weber:2004kj,PrakashReview}).

One possible connection is via oscillations of the compact star,
which on the one hand are affected by the transport properties
of the interior, and on the other hand may have observable
effects on the behavior of the star. The bulk viscosity is
one of the relevant transport properties, and is expected to
play an important role in suppressing both vibrational and rotational
oscillations. One particularly interesting application
involves $r$-modes
\cite{Friedman:2001as,Andersson:2002ch,Kokkotas:2001ze,Madsen:1999ci}.
If the viscosity of the star
is too low, unstable $r$-mode bulk flows will take place, which
quickly spin the star down, removing its angular momentum as gravitational
radiation. The fact that we see quickly-spinning compact stars (millisecond
pulsars) puts limits on the internal viscosity. If we can calculate
the viscosity of the various phases of quark matter then the observations
can be used to rule out some of those phases.
There are various complications to this simple picture. Viscosity is
very temperature-dependent, so to obtain useful limits
we need good measurements of the temperatures
of these stars. There are also some uncertainties about additional sources
of damping that could help to quash $r$-modes. But the essential point is
that viscosity calculations of the various phases are of great
potential phenomenological importance, and in this
paper we report on the results of such a calculation.

The phase that we choose to study is the CFL phase of quark matter.
(The bulk viscosity for unpaired, non-interacting quark matter has
been calculated previously \cite{Madsen:1992sx,Wang:1985tg}.) 
It is known that the mass of the strange quark induces a stress
on the CFL phase that may lead to neutral kaon condensation 
\cite{BedaqueSchaefer,Kaplan:2001qk}, producing a ``CFL-$K^0$''
phase. It is not known whether such condensation occurs at
phenomenologically interesting densities, because of large uncertainties
about instanton effects \cite{Schafer:2002ty},
and in this paper we will assume that kaon condensation has not occurred: our
results are only applicable to the CFL phase, where there is a
thermal population of $K^0$ and other mesons, but no condensation.

Bulk viscosity arises from a lag in the response of the system to an
externally-imposed compression-rarefaction cycle. If there are some
degrees of freedom that equilibrate on the same timescale as the
period of the cycle, then the response will be out of phase with
the applied compression, and work will be done. For astrophysical
applications, such as $r$-modes of compact stars, we are interested
in periods of order 1~ms, which is very long compared
to typical timescales for particle interactions.
In quark matter there is an obvious example of a suitably
slowly-equilibrating quantity: flavor. Flavor is conserved
by the strong and electromagnetic interactions, 
and only equilibrates via weak interactions.

In unpaired quark matter, the lightest degrees of freedom are
the quark excitations around the Fermi surface, and their
flavor-changing weak interactions produce a bulk viscosity 
\cite{Madsen:1992sx}.
However, in the CFL phase the quark excitations are gapped
and their contribution to thermodynamic and transport properties
at temperatures below the gap is irrelevant.
Ignoring the quarks, then, the lightest degrees of freedom in CFL
quark matter are the massless superfluid ``$H$'' modes, the electrons
and neutrinos, the (rotated) photon, and the kaons. Of these, only the
kaons carry flavor, so this paper will focus on the their
contribution to flavor equilibration.
In order to
calculate the bulk viscosity of CFL matter at long timescales such as
1~ms we must therefore calculate the production and decay rates of
thermal kaons. We expect the dominant modes to be ones that involve the $H$,
like $K^0\leftrightarrow H\  H$, and ones that involve the leptons, like
$K^\pm \leftrightarrow e^\pm\ \nu$.

This paper is laid out as follows.  Section~\ref{sec:generalities}
describes how the bulk viscosity is related to the production
and decay rates of the kaons.
Section~\ref{sec:dynamics} describes the basic 
thermodynamics of the system including the kaons and superfluid modes.  
Section~\ref{sec:rates}
describes the overall calculation of the rates of the involved 
processes. Section~\ref{sec:results} presents the results and conclusions.

\section{Relating bulk viscosity to microscopic processes}
\label{sec:generalities}

The bulk viscosity is given by \cite{Madsen:1992sx}
\begin{equation}
\zeta = 
\frac{2 \bar V^2}{\om^2 (\de V)^2} \frac{dE}{dt} \ ,
\label{zeta_def}
\end{equation}
where the system is being driven through a small-amplitude
compression-rarefaction cycle with volume amplitude $\de V$
(see \eqn{epsilons} below) and the driving angular frequency is $\om$.
The average power dissipated per unit volume is
\beq
\frac{dE}{dt} = -\frac{1}{\tau \bar V}\int_0^\tau p(t)\frac{dV}{dt}dt \ ,
\label{dEdt}
\eeq
where $\tau=2\pi/\om$.
We can parameterize the volume oscillation by an amplitude $\de V \ll \bar V$
(chosen to be real by convention), and the resultant
pressure oscillation $p(t)$ by a complex amplitude
$\de p$ which determines its strength and phase:
\beq
\ba{rcl}
V(t) &=& \bar{V} + {\rm Re}(\de V\, e^{i\om t}) \\[1ex]
p(t) &=& \bar{p} + {\rm Re}(\de p\, e^{i\om t})
\ea
\label{epsilons}
\eeq
Substituting these into \eqn{dEdt} and \eqn{zeta_def}, we find that
\beq
\ba{rcl}
\dsp \frac{dE}{dt} &=& 
  \dsp -\frac{1}{2}\dsp \om \,{\rm Im}(\de p)\,\frac{\de V}{\bar V} \\[2ex]
\zeta &=&\dsp -\frac{{\rm Im}(\de p)}{\de V} \frac{\bar V}{\om}
\ea
\label{zeta_general}
\eeq
We therefore expect that ${\rm Im}(\de p)$ will turn out to be negative.
To determine this quantity, we note that
the pressure is a function of the temperature and the chemical potentials.
We assume that heat arising from dissipation is conducted away quickly,
so the whole calculation is performed at constant $T$, and
in order to find $\de p$ we only need to know how the chemical potentials
vary in response to the driving oscillation. 
We expect the bulk viscosity to be most strongly influenced by the lightest
excitations that carry flavor, and for the sake of definiteness 
we will take those to be
kaons. Our analysis could easily be modified to treat the case where
the lightest bosons were pions. At this point we do not have to specify 
whether our kaons are $K^0$ or $K^+$.
The relevant chemical potentials are $\mu_d-\mu_s$ for the
$K^0$ and $\mu_u-\mu_s$ for the $K^+$. For the following generic
analysis we will just write the equilibrating chemical potential
as ``$\mu_K$''.

In thermal equilibrium, the distribution of kaons is determined by their
dispersion relations \eqn{Kdisp} and the Bose-Einstein distribution.
When the kaon population goes slightly out of equilibrium in response to
the applied perturbation, strong interaction processes are still
proceeding quickly: only weak interactions are failing to keep up.
This means that we can always characterize our kaon population by
the flavor chemical potential $\mu_K$, so the kaon 
distribution has the form 
$n_K(p)\propto p^2/(\exp((E_K(p)-\mu_K)/T)-1)$
(see Eq.~\eqn{kaon_distribution}), and nonzero $\mu_K$ indicates deviation
from equilibrium.

We express the variations in the chemical potentials for
quark number and strangeness in terms
of complex amplitudes $\de\mu$, and $\de\mu_K$,
\beq
\ba{rcl}
\mu(t) &=& \bar{\mu} + {\rm Re}(\de \mu \, e^{i\om t}) \ , \\
\mu_K(t) &=&\phantom{\bar{\mu}\, +\,} {\rm Re}(\de\mu_K e^{i\om t}) \ .
\ea
\label{mu_epsilons}
\eeq
Note that the term $-m_s^2/(2\mu)$ which is often 
described as an ``effective chemical potential'' is already included 
in the kaon dispersion relation \eqn{Kdisp},
so in equilibrium, $\mu_K$ is zero.
The pressure amplitude is then
\beq
\de p =
  \frac{\p p}{\p \mu}\Bigr|_{\mu_K} \de \mu
 +\frac{\p p}{\p \mu_K}\Bigr|_{\mu} \de \mu_K
 = n_q \de \mu + n_K  \de \mu_K\ ,
\label{dp_full}
\eeq
(From now on all partial derivatives with respect to $\mu$
will be assumed to be at constant $\mu_K$, and vice versa.)
In principle one might worry that what we have called ``$n_K$'' 
is really $n_d-n_s$ (or $n_u-n_s$), but at temperatures 
below the gap, and in the absence of kaon condensation, 
thermal kaons make the dominant contribution to 
$n_d-n_s$ and $n_u-n_s$.
From \eqn{dp_full} and \eqn{zeta_general} we find
\beq
\zeta = -\frac{1}{\om}\frac{\bar V}{\de V}\Bigl(
 \bar n_q{\rm Im}(\de\mu) + \bar n_K{\rm Im}(\de\mu_K ) \Bigr) \ .
\label{zeta}
\eeq
To obtain the imaginary parts of the chemical potential amplitudes,
we write down the rate of change of the corresponding conserved quantities,
\beq
\ba{rclcl}
\dsp \frac{dn_q}{dt} 
 &=&\dsp \frac{\p n_q}{\p \mu}\frac{d\mu}{dt}
        +\frac{\p n_q}{\p \mu_K}\frac{d\mu_K}{dt}
 &=&\dsp -\frac{n_q}{\bar V} \frac{dV}{dt} \ , \\[2ex]
\dsp \frac{dn_K}{dt} 
 &=&\dsp \frac{\p n_K}{\p\mu}\frac{\p \mu}{dt}
       + \frac{\p n_K}{\p \mu_K}\frac{d\mu_K}{dt}
 &=&\dsp -\frac{n_K}{\bar V} \frac{dV}{dt} - \Ga_{\rm total} \ .
\ea
\label{ndots1}
\eeq
All the partial derivatives are evaluated at equilibrium, $\mu=\bar\mu$ and
$\mu_K=0$.
The right hand term on the first line expresses the fact that quark
number is conserved, so when a volume is compressed, the quark density
rises.  On the second line, which gives the rate of change of kaon
number, there is such a term from the compression of the existing kaon
population, but there is also a net kaon annihilation rate of
$\Gamma_{\rm total}$ kaons per unit volume
per unit time, which
reflects the fact that weak interactions will push the kaon density
towards its equilibrium value. 
The annihilation rate will be calculated from the microscopic
physics in section \ref{sec:rates}.
For small deviations from equilibrium
we expect $\Ga_{\rm total}$ to be linear in $\mu_K$, 
so it is convenient to write the rate in
terms of an average kaon width $\ga_K$, which is defined in terms 
of the total rate by writing
\beq
\Gamma_{\rm total}
= \ga_K\,\frac{\p n_K}{\p \mu_K}\de\mu_K e^{i\om t}
\label{gamma_K}
\eeq
where the derivatives are evaluated at $\mu_K=0$.

We now substitute the assumed oscillations \eqn{epsilons} and
\eqn{mu_epsilons} in to \eqn{ndots1}, and solve to obtain
the amplitudes $\de\mu$ and $\de\mu_K$ in terms of the amplitude $\de V$
and angular frequency $\om$ of the driving oscillation. Inserting their
imaginary parts in \eqn{zeta} we obtain the bulk viscosity

\beq
\zeta = C\frac{ \gaeff }{\om^2 + \gaeff^2}
\label{zeta_K}
\eeq
where
\beq
\ba{rcl}
\gaeff &=&\dsp \ga_K\left(1 -
  \frac{\dsp\Bigl(\deriv{n_K}{\mu}\Bigr)^2}{\dsp\deriv{n_q}{\mu}\deriv{n_K}{\mu_K}}
 \right)^{-1} \approx \ga_K \\[6ex]
C &=&\dsp \Bigl(\deriv{n_q}{\mu}\Bigr)^{-1}\frac{\dsp
 \left( \bar{n}_K \deriv{n_q}{\mu} - \bar{n}_q \deriv{n_K}{\mu} \right)^2}{
  \dsp \deriv{n_K}{\mu_K}\deriv{n_q}{\mu}-\Bigl(\deriv{n_K}{\mu}\Bigr)^2}
\approx 
  \Bigl( \deriv{n_K}{\mu_K}\Bigr)^{-1}
  \left( \bar{n}_K - \bar{n}_q  \deriv{n_K}{\mu}
         \Bigl( \deriv{n_q}{\mu}\Bigr)^{-1} \right)^2
\ea
\label{C_gamma}
\eeq
The approximate forms on the right hand side
are valid for $T\ll\mu$.
They follow
from the fact that
all the derivatives of the kaon free energy go to zero as $T\to 0$, so $n_K$
and its derivatives are suppressed relative to $\bar n_q$ and
$\p n_q/\p\mu$, which are of order $\mu^3$ and $\mu^2$ respectively.

To evaluate $C$ and $\gaeff$
we need the particle densities and their derivatives,
which follow from the full free energy of the system,
$\Om = \Om_{\rm CFL-quarks}(\mu) + \Om_K(\mu,\mu_K)$, where
$\Om_{\rm CFL-quarks}$ is the CFL quark free energy at zero temperature
\cite{Alford:2002kj} and $\Om_K$ is the kaon free energy 
\eqn{kaon_distribution}. Then $n_q$ and $dn_q/d\mu$ come dominantly from 
$\Om_{\rm CFL-quarks}$, and all the other quantities in \eqn{C_gamma}
come from $\Om_K$. We can then see that
the kaon free energy depends on $E_K(p)-\mu_K$, and we choose $m_K$ to be
independent of $\mu$, so from the kaon
dispersion relation \eqn{Kdisp} we see that
the kaon free energy is a function of
$\mu_K + m_s^2/(2\mu)$. This means that
\beq
\deriv{n_K}{\mu} = -\frac{m_s^2}{2\mu^2} \deriv{n_K}{\mu_K}
\eeq
so terms in \eqn{C_gamma} involving $dn_K/d\mu$ are suppressed 
relative to those involving $dn_K/d\mu_K$.

From \eqn{zeta_K} we can already see how the bulk viscosity depends
on the angular frequency $\om$ of the oscillation and the equilibration rate
$\ga_K$. At fixed $\ga_K$, the bulk viscosity decreases as the
oscillation frequency rises; it is roughly constant for
$\om\lesssim\ga_K$, and then drops off quickly as $1/\om^2$ for
$\om\gg\ga_K$.  At fixed $\om$, the bulk viscosity is dominated by
processes with rate $\ga_K\sim\om$, and their contribution is
proportional to $1/\ga_K$. If we imagine varying the rate but keeping
other quantities fixed (e.g.~by varying the coupling constant of the
equilibrating interaction), then for $\ga_K\ll\om$ or $\ga_K\gg\om$
the bulk viscosity tends to zero.  Thus very fast processes, such as
strong interactions, are not an important source of bulk
viscosity. The limit of zero equilibration rate {\em and} zero
frequency is singular, and depends on the order of limits.

In this paper we will also be concerned with the temperature
dependence of the bulk viscosity. This cannot be straightforwardly
read off from \eqn{zeta_K} because the rates and particle densities 
depend on the temperature in complicated ways; however we 
expect that as we go to higher temperatures
($T \gg m_K,\mu_K$) the
bulk viscosity will grow because the kaon density is rapidly increasing.
In the limit of low temperature we expect the viscosity to
be suppressed by $\exp((-m_K+\mu^{\rm eff}_K)/T)$ 
as the thermal kaon population disappears.

\section{Dynamics of the light modes}
\label{sec:dynamics}

In this section we lay out the properties of the lightest modes of the
system, since they will dominate the transport properties.  There is
an exactly massless scalar Goldstone boson associated with spontaneous
breaking of baryon number, and some light pseudoscalars associated
with the spontaneous breaking of the chiral symmetry. We will ignore
the $\eta'$ mode associated with the breaking of $U(1)_A$, since $U(1)_A$
is explicitly broken in QCD at moderate densities.

\subsection{The superfluid ``$H$'' mode}
\label{sec:H}
The CFL quark condensate breaks the exact $U(1)_B$ baryon number symmetry of
the QCD Lagrangian, creating a superfluid with
an exactly massless Goldstone boson $H$.
The Lagrangian for the superfluid mode of the CFL phase is
\cite{Son:2002zn}
\beq
L_{\rm eff} = \frac{N_c N_f}{12\pi^2}
            \bigg[(\p_0\phi - \mu)^2 - (\p_i\phi)^2\bigg]^2
\eeq
This Lagrangian is correct to leading (zeroth) order in $\al_s$ and
to leading order in the derivatives of the $\phi$ field.  This can be 
rescaled to give a conventionally normalized
kinetic term, and the total time-derivative term can be dropped
\cite{Manuel:2004iv}, giving
\beq
L_{\rm eff} = \half(\p_0\phi)^2 - \sixth (\p_i\phi)^2 
   - \frac{\pi}{9\mu^2}\p_0\phi(\p_{\mu}\phi)^2 
   + \frac{\pi^2}{108\mu^4} (\p_{\mu}\phi \p^{\mu}\phi)^2
\eeq
Ignoring the interaction terms for the moment,
the dispersion relation for the $H$ particle is
\beq
E_H(p) = v_H\,p
\eeq
where $v_H^2 = 1/3$ is the ratio of the 
spatial and temporal derivatives above.
In thermal equilibrium at temperature $T$, 
the $H$ bosons have free energy
\beq
\Om_H = \frac{T}{2\pi^2}\int_0^\infty dk\, k^2 
  \ln(1-\exp(-v_H k/T))
      = -\frac{\pi^2}{90 v_H^3} T^4
\eeq
and number density
\beq
n_H = \int \frac{d^3 k}{(2\pi)^3} \frac{1}{\exp(E_H/T)-1} 
    = \frac{\zeta(3)}{\pi^2 v_H^3} T^3 \ .
\eeq

The cubic and higher-order terms allow a single $H$ to decay into multiple
$H$ particles.
For energies far below $\mu$ the dominant process is
$H\to HH$ \cite{Manuel:2004iv,Manuel:2005hu},
whose rate can be calculated by taking the imaginary part of the
1-loop $H$ self energy. Higher order corrections to
the self energy are ignored, both in Ref.~\cite{Manuel:2004iv} and
in this paper.  These corrections could be calculated by taking
into account quantum mechanical interference, the Landau-Pomeranchuk-Migdal 
(LPM) effect.  This should only introduce a difference of $O(1)$ in the
coefficient of the self-energy but would not change the parametric result. 
The self energy is given by Eq.~(3.7) in Ref.~\cite{Manuel:2004iv},
\beq
\Sigma_H(p_0,p) = -\frac{4\pi^2}{81 \mu^4} \sum_{s_1,s_2 = \pm}
   \int \frac{d^3 k}{(2\pi)^3} F(p_0,p,k) \left(\frac{s_1 s_2}{4 E_1 E_2}
  \, \frac{1+f(s_1 E_1) + f(s_2 E_2)}{i\om - s_1 E_1 - s_2 E_2}\right),
\eeq
where
\beq
\ba{c}
F(p_0,p,k) \equiv \bigg[p_0^2 - v^2 p^2 - 2vk(p_0-vk)\bigg]^2, \\[1ex]
f(E) \equiv 1/(e^{E/T}-1),\qquad
E_1 \equiv vk,\qquad  E_2 \equiv v|\bm{p} - \bm{k}|.
\ea
\eeq
Ref.~\cite{Manuel:2004iv} showed that the real part of this self-energy
is parametrically smaller than the imaginary part, so we will only
concern ourselves with the imaginary part, which we will call $\Pi_H$.
There is no contribution to the imaginary part when $s_1 = s_2 = -1$ as
there is no pole in the integral.  One can also show that the 
two terms where the signs of $s_1$ and $s_2$ are opposite are identical.
We can then rewrite this in a slightly simpler and more suggestive
form that will be used in Section \ref{sec:rates}.
\beq
\ba{rcl}
\Pi_H(p_0,p) &=& \dsp \frac{2\pi^3 p_0^2}{81 \mu^4 v^4} \frac{1}{1+f(p_0)} 
 \int \frac{d^3 k}{2 k_0 (2\pi)^3}   F(p_0,p,k) G(p_0,p,k) \ , \\[3ex]
 G(p_0,p,k) &=& \dsp (1+f(E_1)) 
\Bigg(\frac{1+f(E_2)}{E_2}\  \de(p_0 - E_1 - E_2)
   +\ 2\ \frac{f(E_2)}{E_2}\  \de(p_0 - E_1 + E_2)\Bigg)\ .
\ea
\label{PiH}
\eeq
The $H$ propagator can then be written as follows
\beq
D_H(p_0,p) = \frac{1}{p_0^2 - v_H^2p^2 + i\Pi_H(p_0,p)}
\label{H_prop}
\eeq
and we will use this expression in section~\ref{sec:rates}.
It will be useful in the calculation of the decay rates to have the 
$H$ self-energy at momenta and energies close to mass shell ($p_0=vp$).
The self-energy is discontinuous at this point so there are
two values $\Pi_H^+$ and $\Pi_H^-$ depending on whether
$p_0$ tends to $vp$ from above or below,
\beq
\ba{rclcl} 
\Pi_H^+(p) &=&\dsp \lim_{\ep\to 0} \Pi_H(vp+\ep,p)
 &=& \dsp-\frac{\pi p}{81 \mu^4 v} \frac{1}{1+f(vp)}
\int_0^p dk\, I(p,k) \ , \\[3ex]
\Pi_H^-(p) &=&\dsp \lim_{\ep\to 0} \Pi_H(vp-\ep,p)
 &=& \dsp\frac{2\pi p}{81 \mu^4 v} \frac{1}{1+f(vp)}
\int_p^\infty dk\, I(p,k) \ .
\ea
\label{Pi_H_pbar}
\eeq
where $I(p,k) = k^2 (p-k)^2 (1+f(vk))f(vk-vp)$.  
As $T\to 0$,  $\Pi_H^+(p)\propto p^6/\mu^4$, and $\Pi_H^-(p)\to 0$.

\subsection{Pions and Kaons}
\label{sec:pi_and_K}
The CFL quark condensate breaks the approximate $SU(3)$ chiral symmetry of the
QCD Lagrangian, creating eight light pseudoscalar
pseudo-Goldstone mesons. This octet
is just a high-density version of the pion/kaon octet. It is described by
an effective theory \cite{Casalbuoni:1999zi,Son:1999cm}
\beq
L_{\rm eff} = \quarter f_\pi^2 \Tr\Bigl(
  \nabla_0\Si \nabla_0\Si^\dagger - v_\pi^2 \p_i\Si \p_i\Si^\dagger\Bigr)
 + \cdots
\label{Sigma}
\eeq
where $\Sigma=\exp(iP^a\lambda^a/f_\pi)$, and the normalization of
the GellMann matrices is $\tr(\la^a\la^b)=2\de^{ab}$, which yields
a conventionally normalized kinetic term for the 
Goldstone boson fields $P^a$.
At asymptotic densities, weak-coupling calculations give \cite{Son:1999cm}
\beq
f_\pi^2 = \frac{21-8\log(2)}{18}\left(\frac{\mu^2}{2\pi^2}\right) 
  \hspace{1cm} v_\pi^2 = \frac{1}{3} \ .
\label{fpi}
\eeq
When weak interactions have equilibrated, 
the pseudoscalars $P=\pi^\pm,K^\pm,K^0,\overline{K^0}$
have dispersion relations \cite{Son:1999cm,BedaqueSchaefer}
\begin{equation}
E_{P} = -\mueff_{P} + \sqrt{v_\pi^2 p^2 + m_{P}^2} \ ,
\label{Kdisp}
\end{equation}
where
\beq
\ba{rcl}
\mueff_{\pi^\pm} &=&\dsp \pm \frac{m_d^2-m_u^2}{2\mu} \ , \\[1ex]
\mueff_{K^\pm} &=&\dsp \pm \frac{m_s^2-m_u^2}{2\mu}\ , \\[1ex]
\mueff_{K^0,\overline{K^0}} &=&\dsp \pm \frac{m_s^2-m_d^2}{2\mu}\ .
\ea
\label{mueff}
\eeq
Because $m_s\gg m_u,m_d$, the $K^0$ and $K^+$ are expected to
have the smallest energy gap, and so we focus on their contribution
to the bulk viscosity.
We are interested in studying small departures from equilibrium, where
each meson has an additional chemical potential $\de\mu_P$.
It will turn out that the $K^0$ makes the dominant contribution,
so only $\de\mu_{K^0}$ is relevant (Sec.~\ref{sec:K0_rates}).

The expression for the bulk viscosity \eqn{zeta_K} contains terms
of the form $\p n_K/\p\mu$, which take into account the fact that
the meson distributions depend on the meson dispersion relations, which via 
\eqn{mueff} depend on the quark chemical potential.
In this paper we treat the meson masses $m_{K^0}$ etc as constants,
but in perturbative calculations they also depend on $\mu$ and
the CFL pairing gap $\De$ (see section~\ref{sec:meson_mass}).

In thermal equilibrium at temperature $T$ and with chemical potential $\mu_P$, 
the free energy and number density of a meson $P$ is
\begin{eqnarray}
\Om_{P} &=& \frac{T}{2\pi^2}\int_0^\infty dk\, k^2\, 
    \ln\bigl(1-\exp(-(E_{P}-\de\mu_{P})/T)\bigr) \\
n_{P} &=& -\frac{\p \Om_{P}}{\p \mu_{P}} 
    = \frac{1}{2\pi^2} \int_0^\infty dk\ k^2 
      \frac{1}{\exp((E_{P}-\de\mu_{P})/T)-1}
\label{kaon_distribution}
\end{eqnarray}
When weak interactions have equilibrated $\de\mu_P=0$, but when
weak interactions are out of equilibrium the mesons may have nonzero
chemical potentials.

\subsection{Pseudo-Goldstone-boson masses}
\label{sec:meson_mass}
Although we treat the masses as constants, they are predicted to have
density dependence in high density QCD \cite{Schafer:2002ty}
\beq
\ba{rcl}
m_{\pi^\pm}^2 &=&\dsp \frac{1}{f_\pi^2} (2A + 4Bm_s)(m_u+m_d) \ , \\[2ex]
m_{K^\pm}^2 &=&\dsp \frac{1}{f_\pi^2} (2A + 4Bm_d)(m_u+m_s) \ , \\[2ex]
m_{K^0,\overline{K^0}}^2 &=&\dsp \frac{1}{f_\pi^2} (2A + 4Bm_u)(m_d+m_s) \ ,
\ea
\label{meson_mass}
\eeq
where $A$ is positive and related to instantons 
\cite{Schafer:1999fe,Manuel:2000wm}. In the limit of asymptotically large 
density the coefficient $A$ can be computed reliably, but at moderate 
density its value is quite uncertain \cite{Schafer:2002ty}.
Asymptotic-density QCD calculations \cite{Son:1999cm,BedaqueSchaefer} 
also yield 
\beq
B =\dsp \frac{3\De^2}{4\pi^2} \ ,
\eeq
although it is not clear how well these expressions can be trusted
at densities of phenomenological interest.
We will assume that $A$ and $B$ are such that there is
no meson condensation at zero temperature, 
which means that all the meson masses
are greater than their effective chemical potentials.

\subsection{Weak interactions between light bosons}
\label{sec:weak_int}

We have argued above that the bulk viscosity will arise from flavor
violation, which will be dominated by conversion between the lightest
pseudo-Goldstone modes (neutral kaons, typically), which carry flavor,
and the superfluid $H$ modes, which are flavorless. The dominant effect 
of the weak interaction will be to introduce mixing between the $K^0$
and the $H$, in the form of a $K^0 \leftrightarrow H$ vertex \eqn{KHmixing}
in the effective theory. 
We now calculate the strength of that coupling.

The Lagrangian density for the $H$ modes can be written in a
nonlinear form analogous to \eqn{Sigma} for the pseudoscalars,
so the leading terms in the CFL effective theory become
\cite{Casalbuoni:1999zi},
\beq
\label{l_cheft}
{\cal L}_{\rm eff} = \frac{f_\pi^2}{4} {\rm Tr}\left[
 \nabla_0\Sigma\nabla_0\Sigma^\dagger - v_\pi^2
 \partial_i\Sigma\partial_i\Sigma^\dagger \right]
+12 f_H^2 \left[
 \nabla_0 Z\nabla_0 Z^* - v^2_H
 \partial_iZ \partial_i Z^* \right]
\eeq
where $Z=\exp(iH/2\sqrt{6}f_H)$ is the 
field related to the breaking of $U(1)_B$ and $f_H$ is the 
corresponding decay constant. 
At large density the coefficients of the CFL effective Lagrangian can 
be determined in perturbation theory. At leading order  $v_\pi^2=v_H^2=1/3$,
$f_\pi$ is given by \eqn{fpi}, and from Ref.~\cite{Son:1999cm} we obtain
\beq
f_H^2 = \frac{3}{4}\left(\frac{\mu^2}{2\pi^2}\right) \ .
\label{f_H}
\eeq

Under an element $(L,R,\exp(i\al))$ of the
the chiral flavor and baryon number symmetry group
$SU(3)_L\times SU(3)_R\times U(1)_V$, the left-handed quarks,
right-handed quarks, and bosons transform as follows:
\beq
\ba{rcl}
q_L &\to& \exp(i\alpha)L\, q_L \ ,\\
q_R &\to& \exp(i\alpha)R \,q_R \ ,\\
\Sigma &\to& L\,\Sigma \, R^{-1} \ ,\\
Z &\to& \exp(i\alpha)\,Z \ .
\ea
\label{transformation}
\eeq
The weak Hamiltonian breaks the approximate flavor symmetry of QCD,
and only acts on left-handed fields.
The elementary process that is relevant to kaon decay is the
conversion between strange quarks and down quarks via
exchange of a $W^\pm$, which can be treated at energy scales
well below 100 GeV as a four-fermion interaction (Ref.~\cite{Donoghue:1992dd}, 
sect.~II-3 and II-4),
\beq
{\cal L}_{\rm weak} = \frac{G_F V_{ud} V_{us}}{\sqrt{2}}
    (\overline s \ga^{\mu} u)_L(\overline u \ga_{\mu} d)_L + h.c.
\label{weak_H}
\eeq
where $V_{ud}V_{us}\approx 0.215(3)$.
In order to determine how this interaction is represented
in the low energy effective theory of the CFL phase,
we introduce the spurion field $\La_{ds}$ which transforms
as $\La_{ds}\to L\La_{ds}L^\dagger$. In the QCD 
vacuum we set  $\La_{ds}=\la_6$ using the usual
notation for the GellMann matrices (Ref.~\cite{Donoghue:1992dd}, sect.~II-2), 
i.e.~$(\La_{ds})_{\alpha\beta}= \de_{2\al}\de_{3\be}+\de_{3\al}\de_{2\be}$. 
Thus each time this spurion field occurs in an
interaction, it mediates a conversion of downness into strangeness, or
vice versa.
The lowest-order terms in the effective theory that involve
such a conversion are obtained by writing down the
lowest-order terms that contain $\La_{ds}$ and are invariant under
spatial rotations and $SU(3)_L\times SU(3)_R\times U(1)_V$:
\beq
\label{cfl_wk}
{\cal L}_{\rm weak} = 
  f_\pi^2 f_H^2 G_{ds} {\rm Tr}\left[\La_{ds} \Bigl(
            \Si\p_0\Si^\dag\, Z\p_0 Z^* 
- v_{ds}^4  \Si\p_i\Si^\dag\, Z\p_i Z^* \Bigr)\right] 
\eeq
where $G_{ds}$ and $v_{ds}$ are new couplings in the effective action.

\begin{figure}[htb]
\begin{center}
\includegraphics[width=10cm]{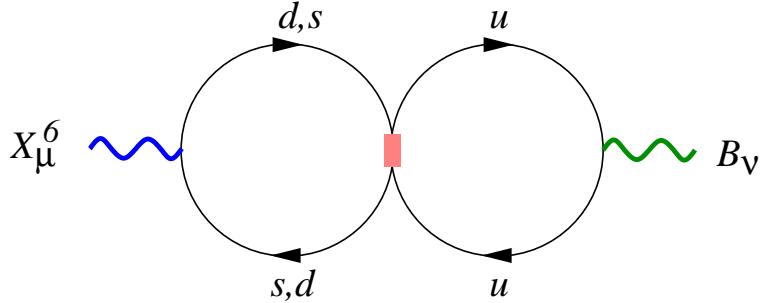}
\caption{Leading contribution to the $X_\mu^6 B_\nu$ 
polarization function in the microscopic theory with gauged
chiral and baryon number symmetries. The shaded bar corresponds
to the vertex of \eqn{weak_H}, which is the low energy limit of a
$W$-boson-mediated interaction.}
\label{fig:polarization}
\end{center}
\end{figure}

Dimensional analysis suggests that $G_{ds}\sim G_F$ and $v_{ds}\sim 1$. 
If the density is large we can be more precise and determine
the coupling constants using a simple matching argument.
For this purpose we gauge the $SU(3)_L$ and $U(1)_B$ symmetries. 
We will denote the corresponding gauge fields $X_\mu^A$ and 
$B_\mu$. The flavor-violating term \eqn{cfl_wk} in the effective
action leads to mixing between the $X_\mu^6$ and $B_\mu$ gauge
bosons. By matching to a calculation of the mixing term in the
microscopic theory \eqn{weak_H}, we now proceed to
determine $G_{ds}$ in terms of $G_F$.

In the effective theory, the $\Si$ field has one left-handed quark index
so $\p_\mu\Si \to (\p_\mu + X^A_\mu\la_A)\Si$.
For the superfluid mode, $\p_\mu Z \to (\p_\mu + B_\mu)Z$. Substituting into
\eqn{cfl_wk} and evaluating in the CFL vacuum ($\Sigma=1$)
we find the mixing term is
\beq 
{\cal L}_{\rm mix} =  2G_{ds}f_\pi^2 f_H^2 (X_0^6 B_0-v_{ds}^4 X_i^6B_i) \ .
\label{mixing_micro}
\eeq
In the microscopic theory, the corresponding calculation is the
computation of the
$X_\mu^6 B_\nu$ polarization function.
At weak coupling the dominant contribution comes from the 
two-loop diagram shown in Fig.~\ref{fig:polarization}. The evaluation of
the Feynman diagram is described in Appendix~\ref{weak_matching}.
The result \eqn{G_ds} is
\beq
\label{g_match}
G_{ds}= \sqrt{2} V_{ud}V_{us}\, G_F \hspace{1cm} v_{ds}^2 = v^2 = 1/3.
\eeq
It is now straightforward to read off the $K^0\to H$ amplitude. 
Linearizing \eqn{cfl_wk}, we find
\beq
{\cal L} = G_{ds} f_\pi f_H \left( \partial_0 K^0 \partial_0 H 
 - v_{ds}^4 \partial_i K^0 \partial_i H \right)
\label{KHmixing}
\eeq
with $G_{ds}$ and $v_{ds}$ given in (\ref{g_match}).
This leads to a vertex factor for the $K$-$H$ interaction given by
\beq
A = G_{ds} f_\pi f_H (p_0^2 - v_{ds}^4\, p^2);
\label{KH_amp}
\eeq
This is the value of the $K^0$-$H$ vertex in Feynman diagrams
such as Fig.~\ref{fig:K_decay}.
Combining this vertex factor and the Lagrangian for the $H$, we can 
calculate the matrix element for conversion between a kaon with
4-momentum $p$ and two $H$s with 4-momenta $k$ and $q$,
\beq
{\it M}^2_{K^0HH}(p,k,q) =
  \frac{G_{ds}^2f_\pi^2 \ (p_0^2 - v_{ds}^4\, p^2)^2}{144 f_H^2}
  {\bigg(p_0(k\cdot q) + k_0 (p \cdot q) + q_0 (p \cdot k)\bigg)^2}
  |D_H(p_0,p)|^2 \ ,
\label{KHH_amp}
\eeq
where $D_H$ is the $H$-propagator \eqn{H_prop}.

\section{Rates of strangeness re-equilibration processes}
\label{sec:rates}

\subsection{Neutral Kaon Rates}
\label{sec:K0_rates}

\begin{figure}[t]
\begin{center}
\includegraphics[width=0.5\textwidth]{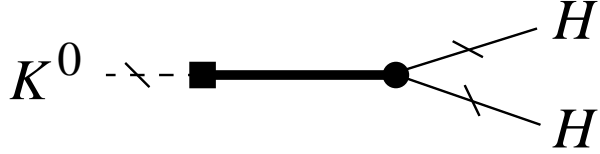}
\caption{The $K^0\to H\ H$ diagram, including the $K^0\to H$
vertex (square), a full $H$ propagator (thick line) and
the $H \to H\ H$ vertex (round). All external lines are amputated.}
\label{fig:K_decay}
\end{center}
\end{figure}
In principle, the correct way to calculate the $K^0$ annihilation rate
is as follows. As noted above, the weak interaction introduces a small
mixing between the $K^0$ and the $H$.  We rediagonalize the kinetic
terms in the effective action, in terms of new fields $E_K$ (which is
the kaon with a tiny admixture of $H$) and $E_H$ (which is the $H$
with a tiny admixture of $K^0$).  We have already seen that the $H$
has a width, arising from the possibility of $H\to H\
H$\footnote{There are also decays involving three or more $H$
particles, but we expect these to be suppressed, since the $H$ is
derivatively coupled, and the greater the number of $H$ particles
involved, the smaller the momentum carried by each of them.}.
When we diagonalize, this will induce
a width for the $E_K$. Because the $E_K$ is almost the same state
as the kaon, the $E_K$ width is a very good estimate of the $K^0$ width.
In terms of the original basis, this width
arises from the vertex shown in Fig.~\ref{fig:K_decay}.
In the interests of brevity, we do not rediagonalize, but
simply calculate the contribution
of the vertex shown in Fig.~\ref{fig:K_decay}
to the kaon annihilation rate. This comes via the processes
$K^0\leftrightarrow H\ H$ and $H\ K^0 \leftrightarrow H$. 
The net kaon annihilation rate is
\beq
\Ga_{\rm total} = \Ga_{\rm forward} - \Ga_{\rm backward}
  = (1-e^{-\de\mu_{K^0}/T}) \Ga_{\rm forward}\ 
    \approx \frac{\de\mu_{K^0}}{T} \Ga_{\rm forward}\ ,
\eeq
where we have used the properties of the Bose-Einstein distributions.
We keep only first order in $\de\mu_{K^0}$, and obtain
the average kaon width $\ga_K$ \eqn{gamma_K}, remembering that
$\dmu_K$ in section \ref{sec:generalities} was the amplitude of a complex
oscillation, so $\de\mu_{K^0} = \dmu_K\exp(i\om t)$,
\beq
\ga_K = \left(\frac{\p n_K}{\p \mu_K}\right)^{-1} 
  \frac{\Ga_{\rm forward}(\de\mu_K = 0)}{T} \ .
\label{gamma_micro}
\eeq
We can therefore obtain the average kaon width simply from the forward
rates $K^0\to H\ H$ and $H\ K^0 \to H$.
The contribution from $K^0\to H\ H $ is
\beq
\Ga_{K^0 \to H H} = \half \int_p \int_{q_1} \int_{q_2}\, |M|^2\, (2\pi)^3\, 
   \de(\bm{p} - \bm{q_1} - \bm{q_2})\, (2\pi)\, \de(p_0 - vq_1 - vq_2)\,
   F_{BE}(p_0,q_1,q_2)
\eeq
with $|M|^2$ given in \eqn{KHH_amp} and $F_{BE}(p_0,q_1,q_2) =
f(p_0 - \de\mu_{K^0})(1+f(vq_1))(1+f(vq_2))$.
The rate for $K^0\ H \to H$ can be obtained by multiplying by $2$ for
the symmetry factor difference, 
switching $q_2 \to -q_2$ in both delta functions and turning 
$(1+f_{vq_2}) \to f_{vq_2}$ to make that $H$ an incoming particle. 
Adding the two contributions, and performing
the $q_2$ integral using the momentum-conserving delta-function,
we obtain the total forward rate
\beq
\Ga_{\rm forward} = G_{ds}^2 f_\pi^2 f_H^2 \int_p\, f(E_K)\, (1+f(E_K))\, 
\frac{(E_K^2 - v^4\, p^2)^2\, \Pi_H(E_K,p)}{
(E_K^2 - v^2\, p^2)^2 + \Pi_H(E_K,p)^2}
\label{K_decay_rate}
\eeq
where $\Pi_H(E_K,p)$ was defined in Eq. \eqn{PiH}.
\begin{figure}[htb]
\begin{center}
\includegraphics[width=10cm]{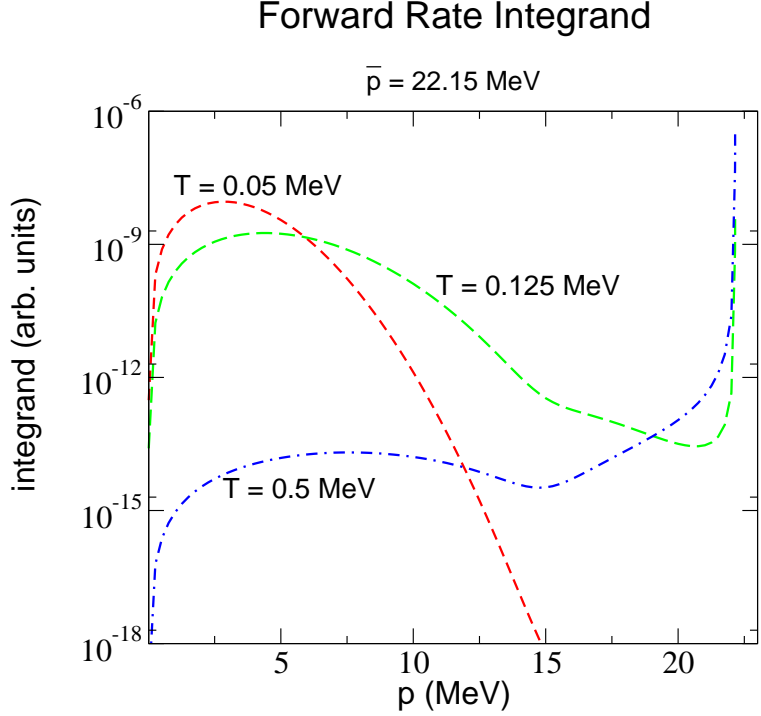}
\caption{Plot of the integrand of \eqn{K_decay_rate}
as a function of momentum for $p<\pbar$. In this plot $\mu=400~\MeV$,
$m_s=120~\MeV$, and $m_K=27.92~\MeV$, so $\pbar = 22.15~\MeV$ and $T_a 
\sim 0.14~\MeV$.
For $T\ll\pbar$ the integral is dominated by low-momentum kaons,
i.e.~$p\ll\pbar$ ($T=0.05~\MeV$ line in plot). 
But as $T$ gets closer to $\pbar$,
the near-singularity at $p=\pbar$ becomes more important, and
at $T=0.5~\MeV$ the integral is actually dominated by the region
where $p$ is very close to $\pbar$.
\label{fig:integrand}
}
\end{center}
\end{figure}

In general this can be evaluated numerically and then combined with 
\eqn{gamma_micro},\eqn{zeta_K},\eqn{C_gamma} to obtain the bulk viscosity.
However for certain temperatures, the integrand is dominated by
momentum that make the denominator as small as possible, i.e. 
when $E_K = v\,p$. 
This corresponds to the $H\leftrightarrow K^0$ resonance, at which
the kaon has a special momentum
$\bar p$ such that an $H$ with that momentum is also on shell,
\beq
\pbar = \frac{m_K^2 - \mueffKz^2}{2\, v \, \mueffKz} = \frac{\dm}{v}\left(1 + \frac{\dm}{2 \mueffKz}\right) \ .
\label{pbar}
\eeq
where $\dm = m_K - \mueffKz$.
Near this momentum, the virtual $H$ in Fig.~\ref{fig:K_decay} is
almost on shell, so momenta close to $\bar p$ dominate the integral.
As long as the numerator is slowly varying, we can approximate the
sharp peak as a delta-function, obtaining
\beq
\Ga_{\rm forward} \approx \frac{G_{ds}^2 f_\pi^2 f_H^2}{18\sqrt{3}\pi} 
  (1+m_K^2/\mueffKz^2) \bar{p}^4 
   \frac{e^{v\bar{p}/T}}{(e^{v\bar{p}/T}-1)^2} \ .
\label{rate_onshell}
\eeq
This expression becomes invalid at very low temperature $T \ll T_a$, 
when there are very few kaons with momentum $\pbar$ so the main
contribution does not come from $p\approx \pbar$, and at very high temperature
$T\gg T_b$ when there are so many thermal kaons with $p>\pbar$ that
they outweigh the contribution from the resonance. One determines
$T_a$ and $T_b$ by setting the non-resonant contribution
equal to the resonant
value from Eq.~\eqn{rate_onshell}, giving
$T_{a,b}$ as a function of $\dm$.
$T_b$ has to be determined numerically and we found that $T_b \sim 9~\MeV$ for
$\dm = 0.1~\MeV$ and monotonically increases as $\dm$ increases, so
$T_b$ is almost always higher than the temperature range that is of
physical interest. $T_a$ is given by the following condition
\beq
T_a \approx \frac{\dm^2}{2\mueffKz}\left(
 \ln\Bigl(\frac{\mu^4}{\dm\, (m_K\, T_a)^{3/2}}\Bigr)\right)^{-1} \ ,
\label{temp_a}
\eeq
where the $T_a$ dependence on the right side is logarithmically weak,
so we expect that $T_a \lesssim \dm^2/(2\mueffKz)$.

The integrand of \eqn{K_decay_rate} is plotted in Fig.~\ref{fig:integrand},
for several different values of the temperature, showing that
for lower temperatures, $T < T_a$, the integrand is a smooth
function, with a broad peak in the low-momentum region, but
as the temperature rises, and the number of thermal
kaon with momentum $\bar p$ rises, the integrand develops a very
sharp peak at $p=\bar p$, where the intermediate $H$ is on shell.

\subsection{Charged Kaon Rates}
\label{sec:K+_rates}

In principle there is also a contribution to kaon number violation
from the charged kaon modes, which necessarily involve charged leptons.
We will now show that this can be neglected compared to the contribution
from the neutral kaons.
The lightest charged kaon is the $K^+$, and the relevant creation/annihilation
reactions are
\begin{eqnarray}
K^+ \leftrightarrow e^+ \nu_e \\
K^+ + e^- \leftrightarrow \nu_e \\
K^+ + \bar{\nu_e} \leftrightarrow e^+. 
\end{eqnarray}
The matrix element for this process has been calculated in 
Ref.~\cite{Jaikumar:2002, Reddy:2003} and is given by
\begin{equation}
A = G f_{\pi} \sin \th_c p_{\mu} \bar{e}(k_1)\ga^{\mu}(1-\ga_5)\nu(k_2),
\end{equation}
where $G$ is the appropriate coupling constant for the charged kaons in the 
medium, which we expect is of order $G_F$.
Summing over initial spins and averaging over final spins, we find
\beq
M^2 = G^2\, f_\pi^2\, \sin^2 \th_c\, m_e^2\, (k_1 \cdot k_2)
\eeq
The rate of the first reaction is 
\bea
\Ga &=& \int_0^{x0} dx \int_{y_1}^{y_2} dy x\, y\, \frac{x+y+\mueffKp/T}{x+y}
  (1-z) F(x,y)\\
z &=& \frac{1}{2 v^2 x y}\bigg((x^2+y^2)(1-v^2) + 2(x+y)\mueffKp/T  
      + 2xy + {(\mueffKp^2 - m_{K^+}^2)/T^2}\bigg) \nn
F(x,y) &=& \frac{\de\muKp}{T}\, \frac{e^{x+y}}{(e^{x+y}-1)(e^x+1)(e^y+1)}.
      \nonumber
\eea
where $F(x,y)$ is the product of the distribution
functions for the kaon
and two leptons to lowest order in $\de \muKp$. 
The rates for the second and third reactions, which are identical
for a massless electron, can be derived from a simple 
change of $x \rightarrow -x$ and $y \rightarrow -y$, respectively.  
They can then be evaluated numerically. 
Note that these calculations are done keeping only the 
lowest order term in $m_e$ and setting $\mu_e = 0$. 

We can then compare this rate to the rate for neutral kaon decay and find
that 
\beq
\frac{\Ga_{K^+}}{\Ga_{K^0}} \sim \frac{T^2}{\mu^2},
\label{K+suppression}
\eeq
for $m_{K^0} \approx m_{K^+}$, so the contribution from charged kaons
is suppressed by a factor of $(T/\mu)^2$. This is to be expected, since
the phase space for quarks is of order $\mu^2 T$,
localized near the quark Fermi surface, whereas the phase space for
electrons is of order $T^3$, since in the CFL phase there is no 
Fermi sea of electrons.

\section{Results}
\label{sec:results}

Our result for contribution
of kaons to the bulk viscosity of CFL quark matter is given by
equations \eqn{zeta_K}, \eqn{C_gamma}, \eqn{gamma_micro}, 
\eqn{K_decay_rate}.
The bulk viscosity is most sensitive to the temperature, and
to the kaon energy gap
\beq
\dm \equiv  m_K - \mueffKz = m_K - \frac{m_s^2-m_d^2}{2\mu} \ .
\label{energy_gap}
\eeq
It also depends on the orthogonal combination $m_K+\mueffKz$, 
the quark chemical potential $\mu$ and CFL
pairing gap $\De$.  As discussed in section 
\ref{sec:meson_mass}, we have no reliable way to calculate
$m_K$ in the density range of interest for compact stars, so in
presenting our results we will treat $\dm$ and $T$ as parameters.

Using the definitions of the densities of kaons and quarks, one
can derive the asymptotic versions of $C$  for temperatures
far above and far below the kaon energy gap (Table \ref{tab:asymptotics}).
One can also derive the low temperature version of the rate and
from that, combined with $C$, we can get the low temperature version
of the bulk viscosity (Table \ref{tab:asymptotics2}).
As one would expect, most kaon-related quantities, including the
bulk viscosity, are suppressed by $\exp(-\dm/T)$ at low temperatures.
This is because the energy gap $\dm$ is the minimum energy required to 
create a $K^0$, so the population of thermal kaons is suppressed by a 
Boltzmann factor.

\begin{table}
\def\st{\rule[-2ex]{0em}{5ex}}
\[
\begin{array}{c@{\qquad}c@{\qquad}c}
\hline
    \mbox{quantity} & \multicolumn{2}{c}{\mbox{asymptotic form}} \\[1ex]
          & T \ll \dm & m_K \ll T \ll \mu \\
\hline
\st n_K  &  (m_K T)^{3/2}e^{-\dm/T} & T^3 \\
\hline
\st \deriv{n_K}{\mueffKz} & m_K (m_K T)^{1/2}e^{-\dm/T} & T^2 \\
\hline
\st \deriv{n_K}{\mu} & -\frac{m_K^2}{\mu} (m_K T)^{1/2} e^{-\dm/T} 
    & -\frac{m_K}{\mu} T^2 \\
\hline
\st n_q & \mu^3 & \mu^3 \\
\hline
\st \deriv{n_q}{\mu} & \mu^2 & \mu^2 \\
\hline
\st C & m_K^3 (m_K T)^{1/2} e^{-\dm/T} & T^4\\
\hline
\end{array}
\]
\caption{
Asymptotic forms for the densities and the C parameter. Constant
numerical factors are not shown and it is implicitly assumed that
$T<0.57\De$ so that there is a CFL condensate, even
when $T \gg m_K$. 
}
\label{tab:asymptotics}
\end{table}

\begin{table}
\def\st{\rule[-2ex]{0em}{5ex}}
\[
\begin{array}{c@{\qquad}c@{\qquad}c@{\qquad}c}
\hline
    \mbox{quantity} & \multicolumn{2}{c}{\mbox{approximate form}} \\[1ex]
          & T < T_a(\dm) \ll \dm & T_a(\dm) < T \lesssim \dm \\
\hline
\st \Ga_{\rm forward} & G_F^2\, \sqrt{m_K^3\,T^3}\, \dm^5\, e^{-\dm/T} 
     & G_F^2\,\mu^4\,\pbar^4\,e^{-v\pbar/T}\\
\hline
\st \gaeff & G_F^2\, \dm^5 
     & G_F^2\,\mu^4\,\pbar^4\,(m_K\,T)^{-3/2}\,e^{-\dm^2/(2\,\mueffKz)\,T}\\
\hline
\st \zeta & G_F^2\,\dm^5\,m_K^{7/2}T^{1/2}\,e^{-\dm/T}\,\om^{-2}
     & G_F^2\,\mu^4\pbar^4\,m_K^2\,T^{-1}\,e^{-v\pbar/T}/{(\gaeff^2 + \om^2)}\\
\hline
\end{array}
\]
\caption{
Approximate forms of the bulk viscosity and related quantities,
for small $T$. Constant numerical factors are not shown.
The rate has two separate ranges within the $T\ll\dm$ region:
$T < T_a\ll \dm$ and $T_a < T \ll \dm$, where
$T_a\lesssim \dm^2/(2\,\mueffKz)$ \eqn{temp_a}.
Note that $\pbar$ is related to $\dm$ by \eqn{pbar}. 
The low temperature entry for $\zeta$ is in general proportional
to $(\gaeff^2 + \om^2)^{-1}$ rather than just $\om^{-2}$, but in this
temperature range $\gaeff$ is always much less than
astrophysically relevant frequencies ($\om\gtrsim 1$~Hz).
There is no third column for the higher end of the range of temperatures
that we study in this paper, $T\sim m_K,\,\mueffKz$, because
although we can still use
\eqn{rate_onshell} for $\Ga_{\rm forward}$, there is no simple form for
$\deriv{n_K}{\mueffKz}$ and hence for $\gaeff$ or $\zeta$.
}
\label{tab:asymptotics2}
\end{table}

To illustrate the likely contribution of kaons to the bulk viscosity
of quark matter in compact stars, we now evaluate the bulk viscosity
numerically for a range of $\dm$ and $T$.
Our calculations are performed at $\mu=400~\MeV$. 
We vary $\dm$ by varying $m_K$ with $\mueffKz$ fixed at $17.92~\MeV$,
corresponding to $m_s=120~\MeV$.

Compact stars have internal temperatures
in the MeV range immediately after the supernova, and then cool
to temperatures in the keV range over millennia, so we explore the range
$0.01~\MeV \lesssim T \lesssim 10 ~\MeV$.
Since $m_K$ and
$\mueffKz$ are both expected to be of order tens of MeV
\cite{Schafer:2002ty}, we expect $\dm$ to be generically of the same
order, so we explore the range $0.1~\MeV \lesssim \dm \lesssim 10~\MeV$.

The bulk viscosity is determined by the kaon equilibration rate
$\ga_K$ \eqn{gamma_micro} and the coefficient $C$ \eqn{C_gamma}, so
we plot these quantities separately before plotting the
bulk viscosity.

\begin{figure}[htb]
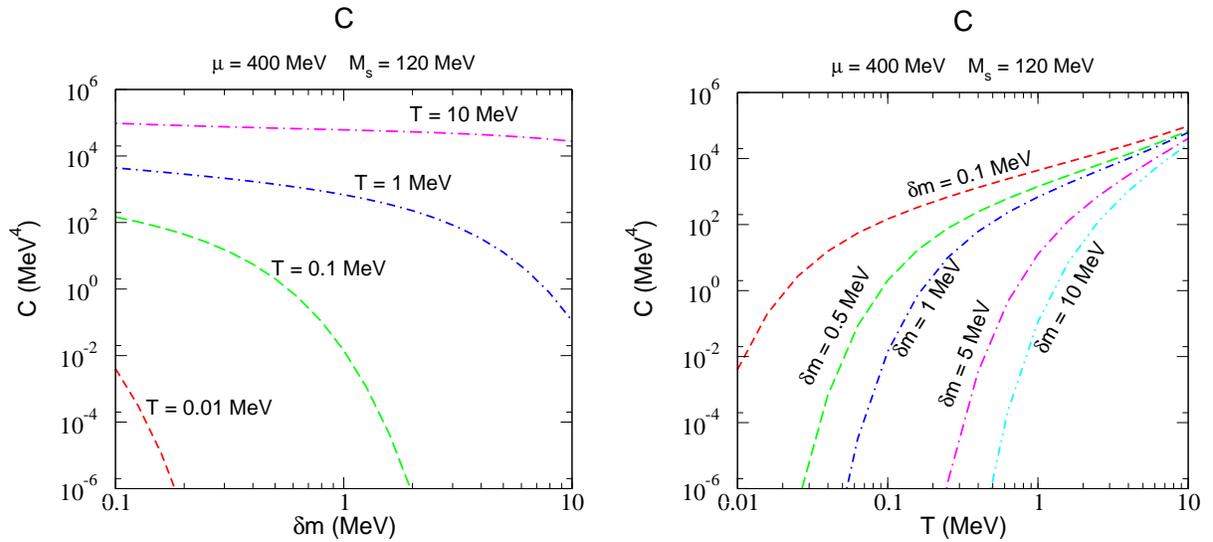

\includegraphics[width=0.49\textwidth]{new_figs/c_mKmuK}
\hspace{0.02\textwidth} 
\includegraphics[width=0.49\textwidth,angle=0]{new_figs/c_varyT}
\caption{Coefficient $C$ \eqn{C_gamma}
as a function of $\dm$ (left panel) and temperature (right panel)}
\label{fig:c}
\end{figure}

\subsection{Proportionality constant $C$ (Fig.~\ref{fig:c})}

In Fig.~\ref{fig:c} we show how $C$ depends on $\dm$ and $T$.
Roughly speaking, $C$ measures how sensitive the kaon and quark number
are to changes in $\mu_K$ and $\mu$.
At low temperatures $T\ll\dm$, $C$ is suppressed
by an exponential factor $\exp(-\dm/T)$, so the curves drop rapidly
in the high $\dm$ region of the left panel, and the low $T$ region
of the right panel.  We also see that the curves for different $\dm$
start to converge at high temperature (right panel). This is because
at high enough temperature (beyond the range that we study)
$C$ would become proportional to $T^4$, independent of $\dm$
(see table \ref{tab:asymptotics}).

\begin{figure}[hbt]
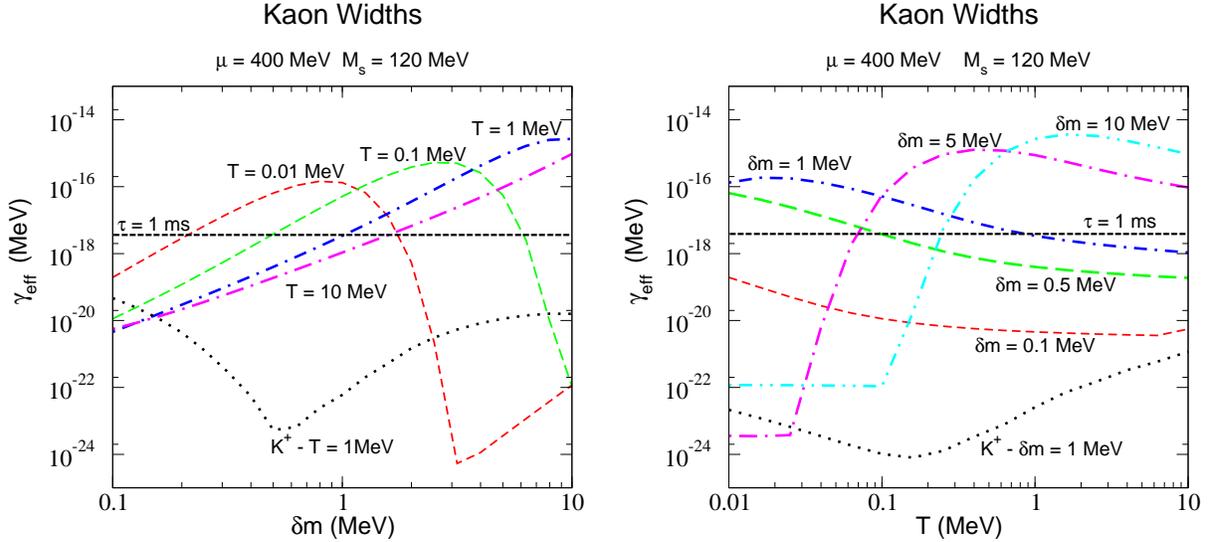

\includegraphics[width=0.49\textwidth]{new_figs/rates_mKmuK}
\hspace{0.02\textwidth} 
\includegraphics[width = 0.49\textwidth]{new_figs/rates_varyT}
\caption{Plot of average $K^0$ decay width $\ga_K$ \eqn{gamma_micro},
\eqn{K_decay_rate} as a function of $\dm$ 
(left panel) and temperature (right panel). The horizontal dashed line 
shows where the width is 1 kHz (${\om/2\pi} = 1~{\rm ms}^{-1}$),
the fastest rotation rate of compact stars. The charged kaon width
is also shown (dotted line) to illustrate that it is a subleading
contribution to strangeness equilibration \eqn{K+suppression}.
The transition that occurs
at $\dm \approx T$ is where the rate becomes dominated by the $H$
resonance (Section \ref{sec:K0_rates}).
}
\label{fig:width}
\end{figure}

\subsection{Kaon width $\gaeff$ (Fig.~\ref{fig:width})}

In Fig.~\ref{fig:width} we show how the neutral kaon effective
width $\ga_K$ depends on $T$ and $\dm$.
(We also show one charged kaon width curve 
to illustrate that it is subleading \eqn{K+suppression}).

The $\dm$-dependence is shown in the left panel.
From Table \ref{tab:asymptotics2}, we expect that at a fixed
temperature $T$, for sufficiently
large $\dm$, $T_a(\dm)$ will become greater than $T$, and
$\ga_K$ will then rise as $\dm^5$. This is seen at the upper end of the
$T = 0.01~\MeV$ curve, and corresponds to the region where
low-momentum ($p<\pbar$) kaons dominate the rate.
For the rest of the $T = 0.01~\MeV$ curve, and
for all the other curves in the plot, the equilibration is dominated by
kaons at the $H$-resonance, with momentum $\pbar$.
The width shows a peak as a function of $\dm$, which follows from the
approximate form for $\gaeff$ given in table \ref{tab:asymptotics2}
(second column). Using \eqn{pbar} to relate $\dm$ to $\pbar$, one
can see $\gaeff \sim \dm^4 \exp(-\dm^2/(2 \mueffKz T))$, which is
peaked at $\dm_{\rm peak} = 2\sqrt{ \mueffKz T}$. For our plots
$ \mueffKz \approx 30~\MeV$, which gives the observed positions
of the peaks.

The $T$-dependence is shown in the right panel.
At the very lowest temperatures $\gaeff$ will have 
a constant value which depends on $\dm$ (Table \ref{tab:asymptotics2}).
This is clear in the curves for $\dm=5~\MeV$ and $\dm=1~\MeV$.
As with the large $\dm$ region of the left panel, this is where
the low momentum kaons are dominating the rate.
In the intermediate temperature region the width rises quickly
and then peaks and drops off slowly. This comes from competition 
\eqn{gamma_micro} between $\Ga_{\rm forward}/T$, which is
monotonically increasing with $T$ \eqn{rate_onshell},
and $(d{n_K}/d{\mu_K})^{-1}$, which is monotonically decreasing with $T$.
At high enough temperature, The expression \eqn{rate_onshell}
for $\Ga_{\rm forward}/T$ rises as $T$,
while $d{n_K}/d{\mu_K}$ rises more quickly, so the width drops.

At high enough $T$, the curve with  $\dm = 0.1~\MeV$ starts to bend upwards.
This feature actually corresponds to $T\gtrsim T_b$ (from Section 
\ref{sec:K0_rates}) where \eqn{rate_onshell} becomes invalid and 
kaons with high momentum ($p>\pbar$) dominate the rate. For the other
curves, $T_b$ is beyond the range that we study.

\begin{figure}[htb]
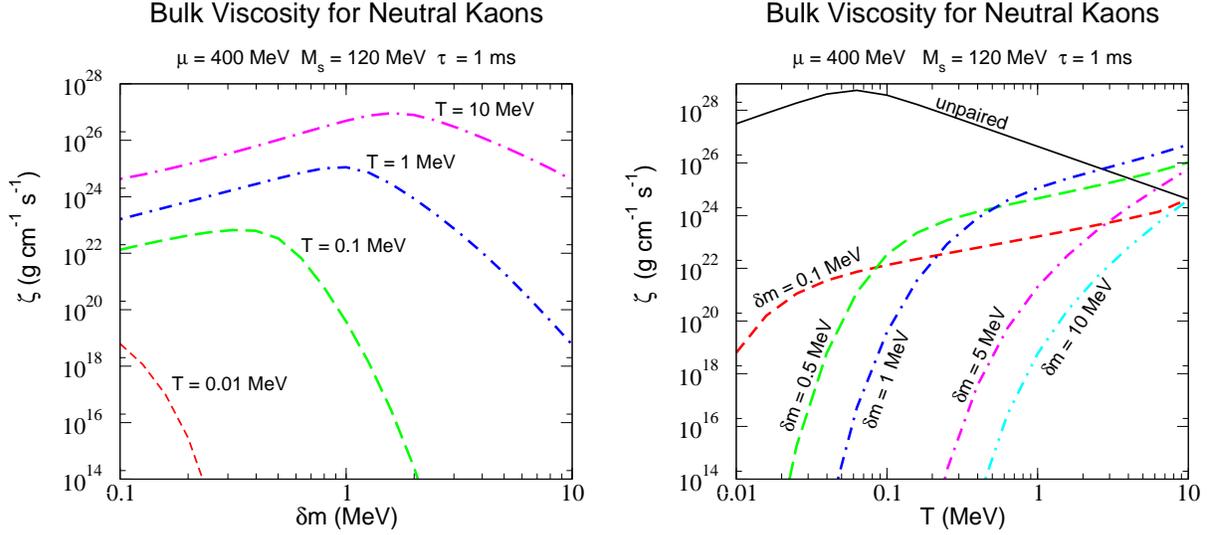

\includegraphics[width=0.49\textwidth]{new_figs/bv_mKmuK}
\hspace{0.02\textwidth}
\includegraphics[width=0.49\textwidth]{new_figs/bv_varyT}
\caption{Plot of bulk viscosity as a function of $\dm$
(left panel) and temperature (right panel).}
\label{fig:bv}
\end{figure}

\begin{figure}[htb]
\bc
\includegraphics[width=0.49\textwidth]{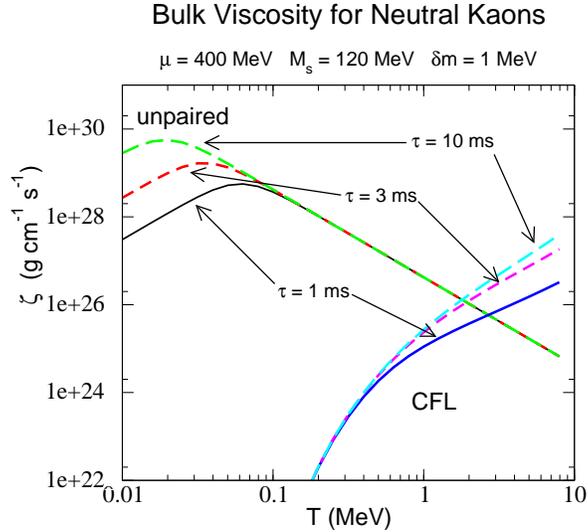}
\ec
\caption{Bulk viscosity as a function of temperature for
oscillations of different frequencies. The curves peaked on the left
are for unpaired 3-flavor quark matter\cite{Madsen:1992sx} 
and the rising curves on the right are our calculation of
the kaonic bulk viscosity of CFL quark matter, for $\dm=1~\MeV$.
}
\label{fig:bv2}
\end{figure}

\subsection{Bulk viscosity $\zeta$ (Fig.~\ref{fig:bv} and \ref{fig:bv2})}
\label{sec:bv}

It is useful to compare the behavior of the bulk viscosity in CFL
quark matter with its behavior in quark matter phases whose
re-equilibration is dominated by ungapped fermionic modes such as the
2SC or single-flavor phases (see solid black line in
Fig.~\ref{fig:bv}).  In such phases, the bulk viscosity shows a peak
as a function of $T$, because $\gaeff$ varies monotonically with $T$,
while $C$ is determined by the phase space at the Fermi surface, and
hence is insensitive to $T$ \cite{Madsen:1992sx,Sa'd:2006qv,Alford:2006gy}.  
This produces a single peak when $\gaeff$ is equal to the 
angular frequency $\om$ of 
the applied compression oscillation (see \eqn{zeta_K}).  As we will
describe below, our results show that in the CFL phase, the situation
is more complicated, because $\gaeff$ is no longer a monotonic
function of $T$, and also because $C$ can vary rapidly as the control
parameters $\dm$ and $T$ are varied.

The dependence of the bulk viscosity (at frequency
$\om/2\pi= 1$~kHz) on the kaon energy gap $\dm$
is shown in the left panel of Fig.~\ref{fig:bv}.
From consideration of the factor of $\gaeff/(\gaeff^2+\om^2)$
in \eqn{zeta_K} we would have expected two peaks for $T=0.01~\MeV$
and $0.1~\MeV$, because from Fig.~\ref{fig:width} we see that at these
temperatures $\gaeff$ passes through $\om=1$~kHz at two different
values of $\dm$.
In fact we get one peak, close to the lower value of $\dm$
at which $\gaeff=\om$. The higher peak is washed out by rapid variation
of $C$ with $\dm$, which occurs when $\dm> T$ (see Fig.~\ref{fig:c}).
Even outside the physically relevant range of $\dm$ shown in our
plots, we do not find additional peaks in the bulk viscosity.

The dependence of the bulk viscosity (again at $\om/2\pi = 1$~kHz) on the
temperature $T$ is shown in the right panel of Fig.~\ref{fig:bv}.
It is a monotonically increasing function of $T$ for all values of $\dm$.  
This is because, as is clear from the right panel of Fig.~\ref{fig:c}, 
$C$ varies rapidly with temperature for all physically
relevant values of $\dm$. 
In fact, the temperature-dependence of $C$ dominates the bulk viscosity,
so the right panel of Fig.~\ref{fig:bv} looks similar to the
right panel of Fig.~\ref{fig:c}. 

Finally, Fig.~\ref{fig:bv2} shows a plot of bulk viscosity 
$\zeta$ as a function
of temperature for different oscillation timescales, $\tau=2\pi/\om$. 
We see that for unpaired quark matter,
$\zeta_{\rm unp}$ is independent of $\om$ at high temperatures,
because $\ga_{\rm unp}$ then rises far above $\om$, 
so, by (see \eqn{zeta_K}), $\zeta = C/\ga_{\rm unp}$.
However, $\zeta_{\rm CFL}$ depends more strongly on $\tau$ at high
temperatures, because $\gaeff$ is not much greater than $\om$
at high temperature (Fig.~\ref{fig:width}), so the $\om$-dependence in
\eqn{zeta_K} is not suppressed. The CFL bulk viscosity becomes larger
as the frequency drops.

\section{Conclusions}
\label{sec:conclusions}

We have calculated the contribution of the lightest pseudo-Goldstone
bosons, the neutral kaons, to the bulk viscosity of CFL quark matter.
Our results are given by equations \eqn{zeta_K}, \eqn{C_gamma},
\eqn{gamma_micro}, \eqn{K_decay_rate}, and are displayed for
reasonable parameter choices in Figs.~\ref{fig:bv} and \ref{fig:bv2}.
The bulk viscosity is most sensitive to the temperature, and to the
kaon energy gap $\dm$ \eqn{energy_gap}.  We find that, as one would
expect, the kaonic bulk viscosity falls rapidly when the temperature
drops below the kaon energy gap, since the the kaon population is then
heavily Boltzmann-suppressed. It is clear from the right hand panel of
Fig.~\ref{fig:bv} that once the temperature falls below the 10 MeV
range (which is expected to occur in the first minutes after the
supernova \cite{Burrows:1986me}) 
the bulk viscosity of CFL quark matter at kHz frequencies is
suppressed by many orders of magnitude relative to that of unpaired
quark matter.

It is noticeable that at low temperatures the suppression is less
severe for smaller kaon energy gaps. However, $\dm$ is a poorly-known
parameter of the effective theory of the pseudo-Goldstone bosons. It
is the difference of the kaon mass and the kaon effective chemical
potential, both of which are expected to be roughly of the order of
10~MeV \cite{Schafer:2002ty}, so it is unnatural to assume that $\dm$
is much smaller than an MeV or so.
For astrophysical applications, it is clear that CFL quark matter
can be sharply distinguished from quark matter by its bulk viscosity
(as by many other transport properties) after the very earliest
times in the life of a compact star.  Also, a rapidly vanishing
bulk viscosity as the temperature drops below $10~\MeV$ could be a
potential observable associated with core collapse supernovae.

In general, bulk viscosity arises from re-equilibration in response to
compression. We have calculated the dominant contribution
to the re-equilibration of flavor in quark matter, and we believe that
this is the dominant contribution to the bulk viscosity as a whole
in the range of frequencies that are of astrophysical interest,
namely zero to 1000 Hz. Any other contribution would have to
come from degrees of freedom that equilibrate on a similar timescale,
and the only possibility that we can imagine is the thermalization
of the low-momentum tail of the thermal distribution of $H$ particles.

Our results highlight several interesting questions for future
research. A natural next step would be to extend our calculation to
the kaon-condensed ``CFL-$K^0$'' phase, which corresponds to allowing 
the kaon energy gap to drop to zero.
There are also some technical issues in our calculation of the 
flavor changing rate that remain to be addressed. The graph shown 
in Fig.~\ref{fig:K_decay} includes the width of the $H$ boson due to the 
one-loop thermal self energy. This corresponds to the resummation 
of a class of diagrams with multiple $H$ boson radiation and 
absorption \cite{Manuel:2004iv}. However, since the mean free path 
associated with small angle two-body collisions is of the same order 
of magnitude as the radiation length 
(the mean free path between $H$-bremsstrahlung events)
this approximation is not 
correct. A more complete approach has to take into account 
quantum mechanical inteference, the Landau-Pomeranchuk-Migdal (LPM)
effect, between different diagrams that have the same final 
state \cite{Arnold:2002zm}.
Another relevant improvement would be
to include higher-order corrections to the $H$
dispersion relation \cite{Zarembo:2000pj} which could have
a strong effect on the collinear splitting amplitude for
$H$ particles.
We do not expect that these improvements will affect our results significantly,
but they would change the numerical prefactors in the rate.
Finally, it should be
noted that we treated the kaon mass as a numerical parameter, but it
is expected to be density-dependent, and its $\mu$-dependence
will feed into our expressions for the bulk viscosity. We expect
that this would only weakly affect our results, but we have not
performed an explicit check.

\medskip
\bc {\bf Acknowledgements} \ec

\noindent
We thank Andreas Schmitt and Tanmoy Bhattacharya
for helpful discussions. 
MGA acknowledges the financial support of a short-term fellowship
from the Japan Society for the Promotion of Science and
the hospitality of the Hadronic Theory Group at Tokyo University,
where this work was completed.
MGA and TS thank the Yukawa Institute for Theoretical Physics at 
Kyoto University, whose YKIS2006 workshop on "New Frontiers in QCD" 
provided a valuable forum for discussions.This research was
supported in part by the Offices of Nuclear Physics and High
Energy Physics of the Office of
Science of the U.S.~Department of Energy under contracts
\#DE-FG02-91ER40628,  
\#DE-FG02-05ER41375 (OJI), 
\#DE-FG02-03ER41260, 
W-7405-ENG-36.   

\appendix
\section{Matching calculation of $G_{ds}$ and $v_{ds}$}
\applabel{weak_matching}

 In this appendix we explain the matching calculation of the
coefficients $G_{ds}$ and $v_{ds}$ in the effective lagrangian.  As
discussed in Sect.~\ref{sec:dynamics} these two coefficients can be
extracted from the $B_\mu-X_\mu^6$ polarization function at zero
energy and momentum transfer. The leading order contribution to the
polarization function at very large baryon density is shown in
Fig.~\ref{fig:polarization}. This two-loop diagram in
can be viewed as the product of two
one-loop polarization functions. This can be made more manifest by
Fierz rearranging the weak interaction given in \eqn{weak_H},
\beq
{\cal L} =- \frac{G_FV_{ud}V_{us}}{\sqrt{2}}
    (\bar{s}\gamma_\mu d)_L(\bar{u}\gamma^\mu u)_L.
\eeq
We can now write the $B_\mu$-$X_\mu^6$ polarization function
at $\omega=\vec{q}=0$ as 
\beq 
\Pi_{\mu\nu}^{XB}= - \frac{G_FV_{ud}V_{us}}{\sqrt{2}}
  \Pi_{\mu\alpha}^{(sd)_L}\Pi_{\alpha\nu}^{(uu)_L}, 
\eeq
where $\Pi_{\mu\nu}^{(sd)_L}$ and $\Pi_{\mu\nu}^{(uu)_L}$
are the polarization functions of the currents $(\bar{s}\gamma_\mu 
d)_L$ and $(\bar{u}\gamma^\mu u)_L$ in the limit of vanishing 
energy and momentum transfer. The polarization functions have
the tensor decomposition
\beq 
 \Pi_{\mu\nu} = f^2 \left(\delta_{\mu 0}\delta_{\nu 0} 
  -v^2 \delta_{\mu i}\delta_{\nu i} \right),
\eeq
where $f$ and $v$ can be viewed as the decay constant and 
velocity of a collective mode. The relevant correlation functions
in the CFL phase were first computed by Son and Stephanov
\cite{Son:1999cm}. Using their results we can write 
\beq 
\Pi_{\mu\nu}^{XB}= - \frac{G_FV_{ud}V_{us}}{\sqrt{2}}
 2 f_\pi^2 f_H^2 \left(\delta_{\mu 0}\delta_{\nu 0} 
  - v^4 \delta_{\mu i}\delta_{\nu i} \right),
\eeq
where $f_\pi$ and $f_H$ are the pion and H decay constants, see
equ.~\eqn{fpi},\eqn{f_H}, and $v$ is the Goldstone boson velocity. Comparing 
with \eqn{mixing_micro} we conclude that 
\beq 
G_{ds}=-\sqrt{2}G_F V_{ud}V_{us}, \hspace{1cm}
v^2_{ds} = v^2 = 1/3.
\label{G_ds}
\eeq


\begin{thebibliography}{99}

\bibitem{Reviews}
  K.~Rajagopal and F.~Wilczek,
  hep-ph/0011333.
  %
  M.~G.~Alford,
  Ann.\ Rev.\ Nucl.\ Part.\ Sci.\  {\bf 51} (2001) 131
  [hep-ph/0102047].
  %
  D.~K.~Hong,
  Acta Phys.\ Polon.\ B {\bf 32}, 1253 (2001)
  [hep-ph/0101025].
  %
  D.~H.~Rischke,
  Prog.\ Part.\ Nucl.\ Phys.\  {\bf 52}, 197 (2004)
  [nucl-th/0305030].
  %
  T.~Sch\"afer,
  hep-ph/0304281.
  %
  S.~Reddy,
  Acta Phys.\ Polon.\ B {\bf 33}, 4101 (2002)
  [arXiv:nucl-th/0211045].

\bibitem{CFL}
M.~Alford, K.~Rajagopal and F.~Wilczek,
  Nucl.\ Phys.\  {\bf B537}, 443 (1999)
  \mbox{[hep-ph/9804403]}.

\bibitem{Weber:2004kj}
  F.~Weber,
  Prog.\ Part.\ Nucl.\ Phys.\  {\bf 54}, 193 (2005)
  [arXiv:astro-ph/0407155].

\bibitem{PrakashReview}
  J.~M.~Lattimer and M.~Prakash,
  Astrophys.\ J.\  {\bf 550}, 426 (2001)
  [astro-ph/0002232];

\bibitem{Friedman:2001as}
J.~L.~Friedman and K.~H.~Lockitch,
gr-qc/0102114.

\bibitem{Andersson:2002ch}
  N.~Andersson,
  Class.\ Quant.\ Grav.\  {\bf 20}, R105 (2003)
  [arXiv:astro-ph/0211057].

\bibitem{Kokkotas:2001ze}
  K.~D.~Kokkotas and N.~Andersson,
  arXiv:gr-qc/0109054.

\bibitem{Madsen:1999ci}
  J.~Madsen,
  Phys.\ Rev.\ Lett.\  {\bf 85}, 10 (2000)
  [arXiv:astro-ph/9912418].

\bibitem{Madsen:1992sx}
  J.~Madsen,
  Phys.\ Rev.\ D {\bf 46}, 3290 (1992).

\bibitem{Wang:1985tg}
  Q.~D.~Wang and T.~Lu,
  Phys.\ Lett.\ B {\bf 148}, 211 (1984).

\bibitem{BedaqueSchaefer}
  P.~F.~Bedaque and T.~Sch\"afer,
  Nucl.\ Phys.\ A {\bf 697} (2002) 802
  [hep-ph/0105150].

\bibitem{Kaplan:2001qk}
  D.~B.~Kaplan and S.~Reddy,
  Phys.\ Rev.\ D {\bf 65}, 054042 (2002)
  [arXiv:hep-ph/0107265].

\bibitem{Schafer:2002ty}
  T.~Sch\"afer,
  Phys.\ Rev.\ D {\bf 65}, 094033 (2002)
  [arXiv:hep-ph/0201189].

\bibitem{Alford:2002kj}
  M.~Alford and K.~Rajagopal,
  JHEP {\bf 0206}, 031 (2002)
  [arXiv:hep-ph/0204001].

\bibitem{Son:2002zn}
  D.~T.~Son,
  arXiv:hep-ph/0204199.

\bibitem{Manuel:2004iv}
  C.~Manuel, A.~Dobado and F.~J.~Llanes-Estrada,
  JHEP {\bf 0509}, 076 (2005)
  [arXiv:hep-ph/0406058].

\bibitem{Manuel:2005hu}
  C.~Manuel,
  PoS {\bf JHW2005}, 011 (2006)
  [arXiv:hep-ph/0512054].

\bibitem{Casalbuoni:1999zi}
  R.~Casalbuoni and R.~Gatto,
  Phys.\ Lett.\ B {\bf 469}, 213 (1999)
  [arXiv:hep-ph/9909419].

\bibitem{Son:1999cm}
  D.~T.~Son and M.~A.~Stephanov,
  Phys.\ Rev.\ D {\bf 61}, 074012 (2000)
  [arXiv:hep-ph/9910491];
  Erratum:
  Phys.\ Rev.\ D {\bf 62}, 059902 (2000)
  [arXiv:hep-ph/0004095].

\bibitem{Schafer:1999fe}
  T.~Sch\"afer,
  Nucl.\ Phys.\ B {\bf 575}, 269 (2000)
  [arXiv:hep-ph/9909574].

\bibitem{Manuel:2000wm}
  C.~Manuel and M.~H.~G.~Tytgat,
  Phys.\ Lett.\ B {\bf 479}, 190 (2000)
  [arXiv:hep-ph/0001095].

\bibitem{Donoghue:1992dd}
  J.~F.~Donoghue, E.~Golowich and B.~R.~Holstein,
  Camb.\ Monogr.\ Part.\ Phys.\ Nucl.\ Phys.\ Cosmol.\  {\bf 2}, 1 (1992).

\bibitem{Jaikumar:2002}
  P.~Jaikumar, M.~Prakash and T.~Sch\"afer,
  Phys.\ Rev.\ D {\bf 66}, 063003 (2002)
  [arXiv:astro-ph/0203088.

\bibitem{Reddy:2003}
  S.~Reddy, M.~Sadzikowski and M.~Tachibana,
  Phys.\ Rev.\ D {\bf 68}, 053010 (2003)
  [arXiv:nucl-th/0306015.

\bibitem{Sa'd:2006qv}
  B.~A.~Sa'd, I.~A.~Shovkovy and D.~H.~Rischke,
  arXiv:astro-ph/0607643.

\bibitem{Alford:2006gy}
  M.~G.~Alford and A.~Schmitt,
  J. Phys. G: Nucl. Part. Phys. {\bf 34} 67 (2007)
  [arXiv:nucl-th/0608019].

\bibitem{Burrows:1986me}
  A.~Burrows and J.~M.~Lattimer,
  Astrophys.\ J.\  {\bf 307}, 178 (1986).

\bibitem{Arnold:2002zm}
  P.~Arnold, G.~D.~Moore and L.~G.~Yaffe,
  JHEP {\bf 0301}, 030 (2003)
  [arXiv:hep-ph/0209353].

\bibitem{Zarembo:2000pj}
  K.~Zarembo,
  Phys.\ Rev.\ D {\bf 62}, 054003 (2000)
  [arXiv:hep-ph/0002123].

\end{thebibliography}
\end{document}